\documentclass{ws-ijmpa}

\begin{document}

\markboth{L. Le\'sniak et al.}
{Analysis of $B^{\pm} \rightarrow K^+ K^- K^{\pm}$ decays}

%
\catchline{}{}{}{}{}
%

\title{\bf ANALYSIS OF $B^{\pm} \rightarrow K^+ K^- K^{\pm}$ DECAYS}

\author{\footnotesize L. LE\'SNIAK$^1$, A. FURMAN$^2$, R. KAMI\'NSKI$^1$, P. \.ZENCZYKOWSKI$^1$}
\vspace{0.5cm}

\address{\it \small 
$^1$ Division of Theoretical Physics, The Henryk Niewodnicza\'nski 
Institute of Nuclear Physics, Polish Academy of Sciences, 31-342 Krak\'ow,
 Poland \\
$^2$ ul. Bronowicka 85/26, 30-091 Krak\'ow, Poland\\
 Leonard.Lesniak@ifj.edu.pl}

\maketitle


\begin{abstract}Rare $B^{\pm}$ decays to three charged kaons are analysed. 
The weak decay amplitudes are derived in the QCD 
factorization approach. The strong final state 
interactions between pairs of kaons are described 
using the kaon scalar and vector form factors. The
 scalar form factors at low $K^+K^-$ effective mass 
distributions are constrained by chiral symmetry
 and are related to the coupled channel meson-meson 
amplitudes describing all the transitions between
 three channels consisting of two kaons, two pions
and four pions. The vector form factors are fitted 
to the data on $e^+e^-$ collisions.
   The model results are compared with the Belle and 
BaBar data.

\end{abstract}
\ccode{PACS numbers: 13.25.Hw, 13.75.Lb, 14.40.Nd}

\section{\bf Theoretical model}
We study rare charged B meson decays into three charged kaons. In seach for direct $CP$ violation
the strong interactions between kaons must be included. Partial wave analysis of the decay 
amplitudes helps in investigation of the density distributions of the Dalitz diagrams.
The QCD quasi-twobody factorization approach for a limited range of the effective $K^+ K^-$
masses less than about 1.8 GeV is used.

The amplitude for the $B^-$ decay into three charged kaons reads: 
\begin{equation}
\begin{split}
&\langle K^-(p_1)K^+(p_2)K^-(p_3)|H|B^-\rangle=\dfrac{G_F}{\sqrt{2}}\{\sqrt{\dfrac{1}{2}}\chi
 f_{K}(M^2_B-s_{23})F^{B\to(K^+K^-)_S}_0(m^2_{K})u\\
&\Gamma^{n^*}_2(s_{23})+\dfrac{2B_0}{m_b-m_s} (M^2_B-m^2_{K})F^{B K}_0(s_{23})v\Gamma^{s^*}_2(s_{23})
+4\overrightarrow{p}_1 \cdot \overrightarrow{p}_2[\dfrac{f_K}{f_{\rho}}A^{B\rho}_0(m^2_{K})u \\&F^{K^+K^-}_u(s_{23})
-F^{BK}_1(s_{23})(w_uF^{K^+K^-}_u(s_{23})+w_dF^{K^+K^-}_d(s_{23})+w_sF^{K^+K^-}_s(s_{23}))]\},\\
\end{split}
\end{equation}
where $G_F$ is the Fermi coupling constant, $f_K$ and $f_{\rho}$ are the kaon and $\rho$ meson decay constants,
$M_B$, $m_K$, $m_b$, $m_s$, $m_u$ and $m_d$ are the masses of the $B$ meson, kaon, $b$-quark, strange quark,
up- and down quarks, respectively; $s_{23}$ is the $K^+(p_2)K^-(p_3)$ effective mass squared,
$\overrightarrow{p}_1$ and $\overrightarrow{p}_2$ are the kaon 1 and kaon 2 momenta in the center
of mass system of the kaons 2 and 3. The functions 
$\Gamma^{n}_2$ and $\Gamma^{s}_2$ are the kaon nonstrange and strange scalar form factors,  
$F^{K^+K^-}_u$, $F^{K^+K^-}_d$ and $F^{K^+K^-}_s$  denote three types of the kaon vector form factors,
$F^{B\to(K^+K^-)_S}_0$ is the form factor of the transition from the $B$ meson to the $K^+K^-$ pair in the 
$S$-wave, $\chi$ is the constant related to the decay of the $(K^+K^-)_S$ state into two kaons,
$F^{B K}_0$, $F^{B K}_1$ and $A^{B\rho}_0$ are the $B\to K$ and $B\to\rho$ transition form factors. 
The constant $B_0= m_{\pi}^2/(m_u+m_d)$, $m_{\pi}$ being the pion mass. Other symbols are defined in terms of
 Wilson's coefficients $a_j$ and the products $\Lambda_u=V_{ub}V_{us}^*$, $\Lambda_c=V_{cb}V_{cs}^*$, $V_{ij}$ 
being the CKM quark-mixing matrix elements:
\begin{equation}
 u=\Lambda_u[a_1+a_4^u+a_{10}^u-(a_6^u+a_8^u)r_{\chi}+\Lambda_c[a_4^c+a_{10}^c-(a_6^c+a_8^c)r_{\chi}],
\end{equation}
\begin{equation}
 v=\Lambda_u(-a_6^u+\frac{1}{2}a_8^u)+\Lambda_c(-a_6^c+\frac{1}{2}a_8^c),~~~~~~~~r_{\chi}=\dfrac{2m_K^2}{(m_b+m_u)(m_u+m_s)}
\end{equation}
\begin{equation}
 w_u=\Lambda_u a_2+(\Lambda_u+\Lambda_c)(a_3+a_5+a_7+a_9),~~~w_d=(\Lambda_u+\Lambda_c)[a_3+a_5-\frac{1}{2}(a_7+a_9)],
\end{equation}

\begin{equation}
 w_s=\Lambda_u[a_3+a_4^u+a_5-\frac{1}{2}(a_7+a_9+a_{10}^u)]+\Lambda_c[a_3+a_4^c+a_5-\frac{1}{2}(a_7+a_9+a_{10}^c)].
\end{equation}

The decay amplitude given by Eq. (1) consists of two parts: the $P$-wave part proportional to the product 
$\overrightarrow{p}_1 \cdot \overrightarrow{p}_2$ and the preceding term being the $S$-wave part. In the final state
of the decay reaction we take into account the elastic kaon-kaon $S$-wave interactions and other transitions like
the $K^+K^-$ annihilations into systems consisting of two and four pions. Thus a system of three coupled 
channels: $\pi\pi, \bar KK$ and $4\pi$ (effective $\sigma\sigma$), labelled by 1,2 and 3, is considered.
A set of the 3x3 transition 
amplitudes $T$ is taken from a unitary model developed in Ref. 1. At first the production functions $R_j$ are introduced 
as follows:
\begin{equation}
R_j(E)=(\alpha_j+\tau_jE+\omega_jE^2)/(1+CE^4),~~~~~~E \equiv m_{K^+K^-},
\end{equation}
where $\alpha_j,\tau_j,\omega_j$ and $C$ are constant parameters. Then the three scalar form factors, written in 
the compact matrix form $\Gamma$ are expressed by
\begin{equation}
\Gamma^*=R+TGR,
\end{equation} 
where $G$ is the matrix of the channel Green functions. The coefficients $\alpha_j,\tau_j$ and $\omega_j$
are constrained by the values of the form factors, calculated in the framework of the chiral
model of Ref. 2, using the numerical results of the lattice QCD \cite{Allton}. The parameter $C$, 
which controls the high energy behaviour of $R$, can be fixed while fitting the data of the BaBar \cite{BaBar}
and Belle \cite{Belle} collaborations. The present model of the scalar form factors satisfies the unitarity
conditions.

The three vector form factors $F^{K^+K^-}_q$ for $q=u,d,s$ defined as
 \begin{equation}
<K^+(p_2)K^-(p_3)|\bar q \gamma_{\mu} q |0>=(p_2-p_3)_{\mu}F^{K^+K^-}_q(s_{23})
\end{equation}
enter into the $P-$wave amplitude.
These functions contain contributions from eight vector mesons $\rho(770)$, 
$\rho(1450)$, $\rho(1700)$, $\omega(782)$, $\omega(1420)$, $\omega(1650)$, $\phi(1020)$ and
$\phi(1680)$. Here the model of Ref. 6 has been used in their parametrization.
\newpage 

\begin{figure}[t!]
  \includegraphics[height=.25\textheight]{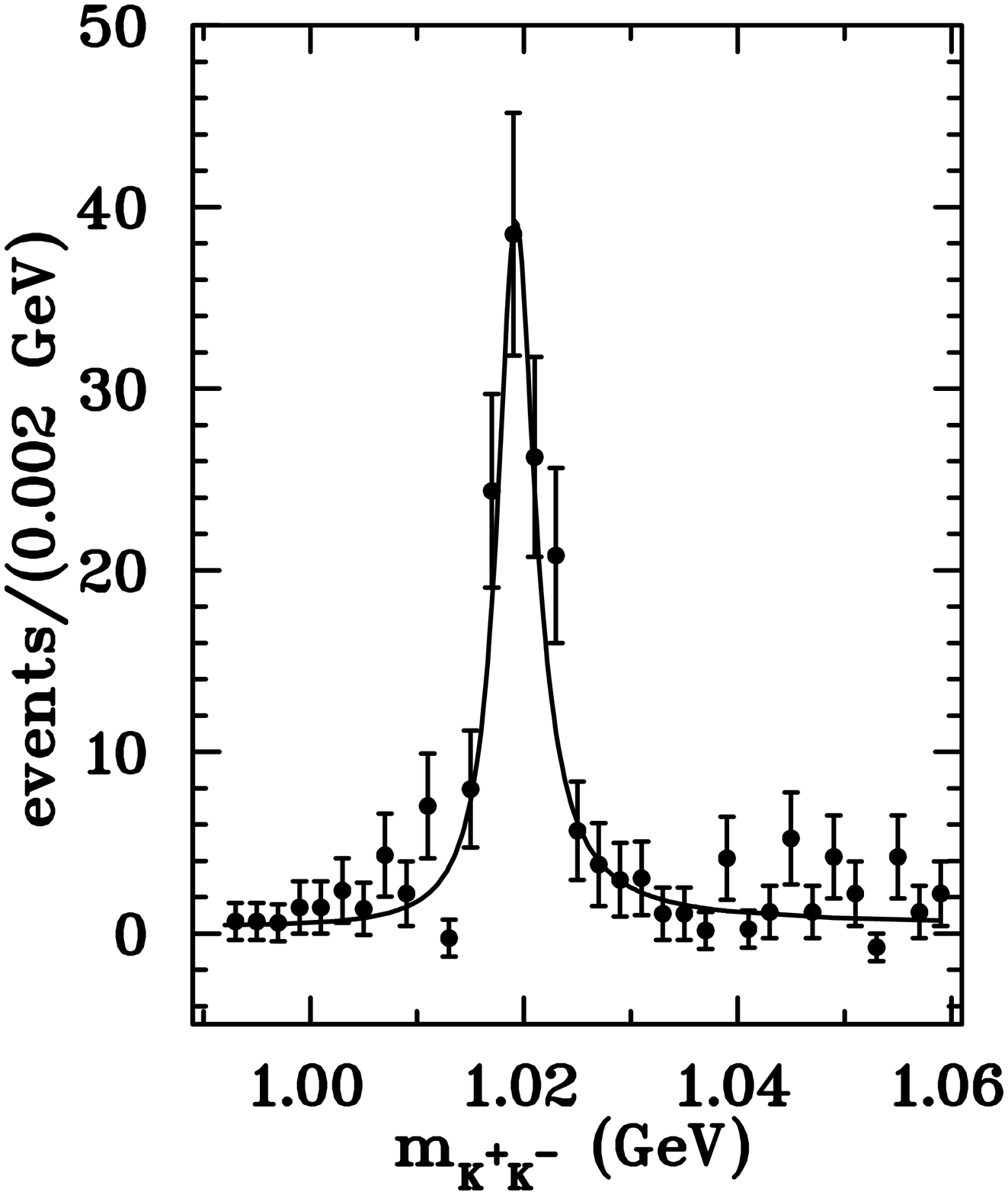}~~ 
  \includegraphics[angle=0,height=.25\textheight]{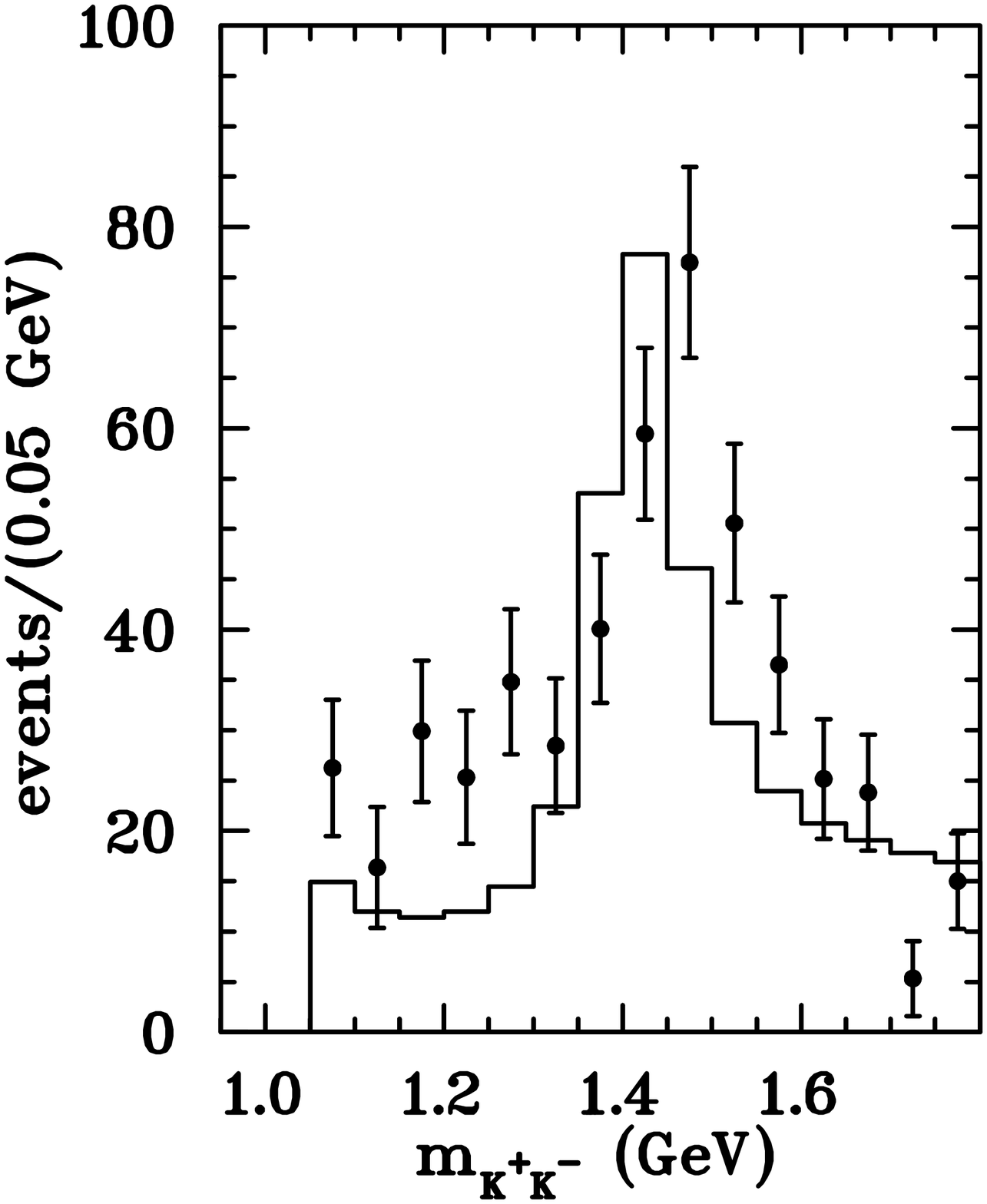}~~
  \includegraphics[height=.25\textheight]{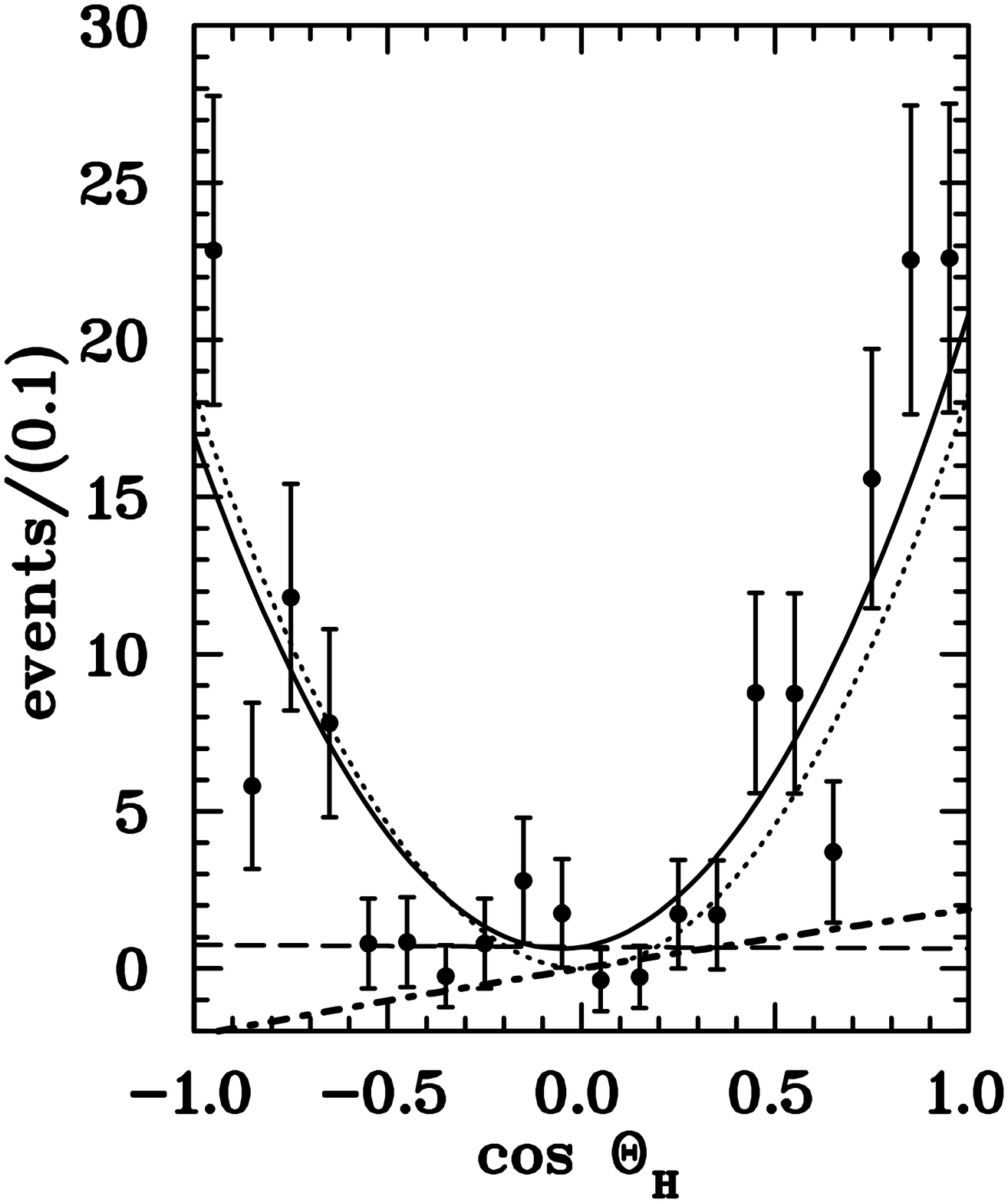} 

\vspace*{6pt}

\caption{The $K^+K^-$ effective mass distributions from the fit to the Belle experimental
data in the $\phi(1020)$ range (a) and between 1.05 GeV and 1.75 GeV (b). 
The theoretical results are shown as the solid line in (a) and as the histogram in (b). 
Helicity angle distributions for events
 in the $K^+K^-$ effective mass up to 1.05 
GeV (c). The dashed line represents the $S$-wave contribution of our model, the dotted line 
- that of the $P$-wave, the dot-dashed - that of the interference term and the
solid line corresponds to the sum of these contributions.}
\end{figure}
\vspace*{-235pt}
\hspace{0.5cm} {\bf a)}\hspace{4cm} {\bf b)}\hspace{4cm} {\bf c)}
\vspace*{230pt}
\vspace*{-1.2cm}
\section{\bf Results}
Preliminary results of the model fits to the data of the 
 Belle \cite{Belle}
Collaboration are shown in Fig. 1. The scalar resonance $f_0(980)$ leads to the threshold enhancement of the $S$-wave amplitude. The 
$K^+K^-$ structure seen near 1.5 GeV can be attributed to another scalar resonance.
Sizable helicity angle asymmetry appears in the $\phi(1020)$ range. 
A potentially large $CP$ asymmetry can be discovered in 
 the mass spectrum dominated by the $S$-wave. However,
new experimental analyses of data with better statistics are needed. The data already
exist! For example, the Belle Collaboration has now five times larger data sample than 
that used in their publication \cite{Belle}. Future results from super B factories will
also be very useful.

\vspace*{-0.8pt}
\section*{Acknowledgments}
This work has been supported in part by the Polish Ministry of Science and
Higher Education (grant No N N202 248135). 

\end{document}